\documentstyle[epsfig,longtable]{aipproc}
\begin{document}
\title{Dark matter distribution in the universe and ultra-high
energy cosmic rays}
\author{Pasquale Blasi$^*$}
\address{$^*$NASA/Fermilab Astrophysics Center}

\maketitle

\begin{abstract}
Two of the greatest mysteries of modern physics are the origin of the 
dark matter in the universe and the nature of the highest energy particles
in the cosmic ray spectrum. We discuss here possible direct and indirect
connections between these two problems, with particular attention to
two cases: in the first we study the local clustering of possible sources 
of ultra-high energy cosmic rays (UHECRs) driven by the local dark
matter overdensity. In the second case we study the possibility that 
UHECRs are directly generated by the decay of weakly unstable super
heavy dark matter. 
\end{abstract}

\section*{Introduction}

Most of our universe is made of a dark elusive component, whose nature
is still unknown. This mysterious component dominates the gravitational
interactions in our world and determines the distribution of visible matter,
that we associate with galaxies, clusters and superclusters of galaxies and
their luminous constituents. We know from the measure of
the values of cosmological parameters, that most of the dark matter 
must be of non 
baryonic nature. This important finding stimulated the blossoming of numerous
proposals of new particles, in the context of theories of Grand Unification
or extensions of the Standard Model.
Some of the possibilities that have been proposed invoke the existence of 
super heavy particles, with mass comparable with the mass of the inflaton.
These particles might be
almost stable and contribute the whole of cold dark matter needed to 
explain cosmological and astrophysical observations. 
Independent of the nature of dark matter, astrophysical observations of 
the visible universe allow us to know how dark matter is distributed in 
space: primordial fluctuations in the density field result today in a 
structured distribution of the dark and visible components in the universe.
Galaxies are part of this clustering, but similar local overdensities can 
be seen on larger scales, up to a few tens Mpc, where the universe starts
being well described by a homogeneous density field. 

An apparently unrelated mystery of modern physics is the existence of
particles with energy in excess of $10^{20}$ eV in the cosmic radiation,
the so-called ultra high energy cosmic rays (UHECRs).  These particles are
thought to be of extragalactic origin, bacause the galactic magnetic field
is unable to confine particles with such energy. As pointed out in the 
pioneering papers by Greisen \cite{greis} and Zatsepin and Kuzmin \cite{kz},
if the sources are distributed homogeneously in the universe, the process
of photopion production off the photons of the cosmic microwave 
background (CMB) should cause a strong suppression in the spectrum 
of UHECRs above $\sim 5\times 10^{19}$ eV, what is now known as
the Greisen-Zatsepin-Kuzmin (GZK) cutoff. Nevertheless we do detect 
particles with energy much in excess of this cutoff energy, and we have been 
unable to identify any straightforward astrophysical source in the direction of
arrival of these events.
Can the two mysteries,  of dark matter and UHECRs be related? We discuss
here two possible connections: in sections 2 and 3 we explore the 
consequences that a local clustering of the sources, driven by 
the corresponding 
dark matter clustering, have on the observed fluxes of UHECRs, in 
comparison with the case of homogeneous
distribution. The initial prediction by Greisen and Zatsepin and 
Kuzmin, was in fact derived in the assumption of sources equally
distributed in space. It was later shown by Berezinsky and Grigorieva 
\cite{beregri79}
that a local overdensity by a factor $\sim 10$ over $\sim 20$ Mpc would
make the problem of the existence of UHECRs less severe. The local 
density of sources can now be extracted from large catalogs, so that a
realistic determination of the density field can finally be used in the
calculation of the fluxes of UHECRs. We use here the PSCz and CfA 
catalogs for the purpose of calculating the density field. The fluxes of 
UHECRs are calculated numerically by montecarlo simulations.

In section 4 we consider a direct connection between dark matter and
UHECRs, in the case that dark matter is composed of super heavy quasi-stable
particles. Particles fulfilling these requirements have been recently 
discussed in \cite{kolb,bere,kuzmin,kt} [see also J. Ellis (these 
proceedings)]. We will review here
the current situation and stress some consequences of this model on the
observable large scale and small scale anisotropies in the arrival directions
of UHECRs.

\section*{Measuring the Galaxy Density Field}

In order to determine the effects of the inhomogeneity of the source
distribution on the cosmic ray spectrum, we need to measure the galaxy
density field, and in particular its dependence on redshift. In general,
it may well be that the sources of UHECRs are of some type not directly
related to other known objects like galaxies, but we assume here that 
the sources of UHECRs have a density field that is proportional to that
of ordinary galaxies. We follow the approach presented in \cite{bbo00}.
The galaxy
density field is usually measured by selecting galaxies from an
imaging sky survey and taking their redshifts. Almost invariably, the
galaxies are selected to be brighter than some limiting flux
$f_{\mathrm{lim}}$ in some band, expressed as an ``apparent
magnitude'' $m_{\mathrm{lim}}=-2.5\log_{10} (f_{\mathrm{lim}}/f_0)$,
where $f_0$ is an arbitrary zero-point. For all (or for some random
subsample) of the galaxies brighter than this, their spectra
are taken and their redshifts $z$ are determined. 

However, we cannot simply use the raw distribution of redshifts from
such a flux-limited survey, regardless of the way the galaxies
were selected. Here we describe the proper way to derive density
fields from galaxy redshift surveys. We limit ourselves to measuring
the density in redshift space, ignoring the effects of deviations from
the Hubble law due to galaxy peculiar velocities. The subject of
galaxy density fields dates back to
Ref. \cite{davis82a}. An educational recent review is that of Ref.
\cite{strauss95a}.

Our results will be based on two surveys. First, we consider the Center for
Astrophysics Redshift Survey (CfA2; \cite{huchra95a}). This survey
comprises about 10,000 galaxy redshifts, selected to be brighter than
$m = 15.5$, (approximately, a $B$-band magnitude). It covers an area
which is about 17\% of the whole sky. However, in order to evaluate
the effects of the density field of galaxies on the cosmic ray
spectrum, we really should probe the density field over nearly the
whole sky. The best sample of galaxies to use for this purpose is the
IRAS PSC$z$ Survey \cite{saunders00a}, which consists of about
15,000 galaxies with infrared fluxes $>0.6$ Jy over about 84\% of the
sky.

A consequence of the flux limits in any survey is that at different
redshifts, a different set of galaxy luminosities $L$ is observed,
determined by the faintest luminosity observable at that redshift
$L_{\mathrm{min}}(z)$. For an Euclidean metric, this luminosity is
related to the flux limit by
$L_{\mathrm{min}}(z) = 4 \pi (H_0 c z)^2 f_{\mathrm{lim}}$.

If the distribution of galaxy
luminosities is described by the galaxy luminosity function $\Phi(L)$,
it is then simple to calculate what fraction of all galaxies is
observable at any redshift:
\begin{equation}
\phi(z) = \frac{\int_{L_{\mathrm{min}}(z)}^\infty dL \Phi(L)}
{\int_{0}^\infty dL \Phi(L)}.
\end{equation}
The quantity $\phi(z)$ is usually referred to in the literature as the
``selection function'' \cite{peebles80a}. The most common methods
used to determine the galaxy luminosity function from the survey
itself are those of \cite{efstathiou88a} and
\cite{sandage79a}. These methods assume that the luminosity function
has universal shape, but {\it not} that galaxies are distributed 
homogeneously.

Figure 1 shows in the top panel
the distribution of galaxies and redshifts in CfA2.  Here we express
galaxy luminosity in terms of the ``absolute magnitude'' $M =
-2.5 \log_{10} L +$ const.
The thick solid line shows the flux limit of the
survey, translated into an absolute magnitude limit at each
redshift. Because of this limit a number of galaxies which are
observable at low redshifts are too faint to be observed at higher
redshift. 

\begin{figure}[thb]
 \begin{center}
  \mbox{\epsfig{file=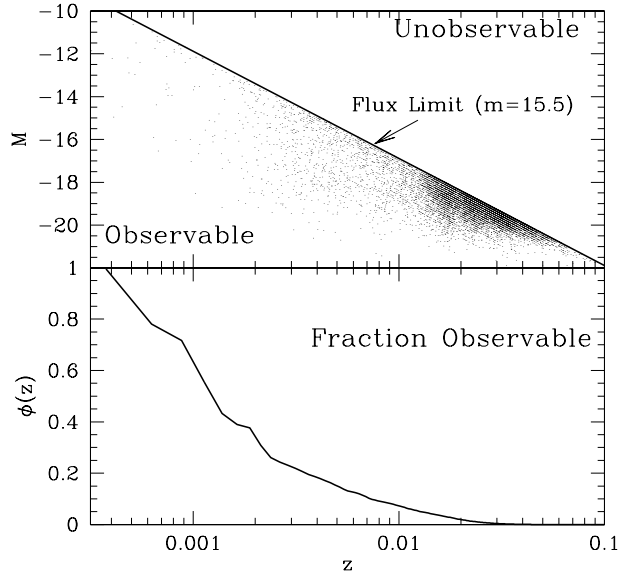,width=9cm}}
  \caption{\em {
The top panel shows the absolute magnitudes
(related to luminosity by $M = -2.5 \log_{10} L +$ const) and
redshifts of CfA2 galaxies. Shown as the thick solid line is the
flux limit, converted to the appropriate absolute magnitude at each
redshift. The bottom panel shows the fraction of galaxies in the range
$-22<M<-10$ that we estimate to be brighter than the flux limit. This
function falls rapidly with redshift. When interpreting the top plot,
remember that the volume probed at low redshift is far smaller than
that probed at high redshift.
}}
 \end{center} 
\label{Mz}
\end{figure}

The fraction of galaxies $\phi(z)$ between absolute
magnitudes $-22<M<-10$ which are unobservable at each redshift is
shown in the bottom panel of Figure
1, based on a fit to the luminosity function in the survey
using the method of \cite{efstathiou88a}. Because the function falls
rapidly from unity, it is clear that even at low redshifts the effects
of the flux limit are important. 
We can use $\phi(z)$ to calculate the expected distribution of
observed galaxies with redshift.
The top panel of Figure 2
compares these expected counts (dotted line) in redshift shells of
thickness 0.001 to the observed counts (solid line) in CfA2. It
appears that locally we are in an overdensity of galaxies of about a
factor of two; note that at large distances, where each shell
corresponds to a considerable amount of volume, the number of 
galaxies is very nearly the expected number. 

\begin{figure}[thb]
\label{cfapscz}
 \begin{center}
  \mbox{\epsfig{file=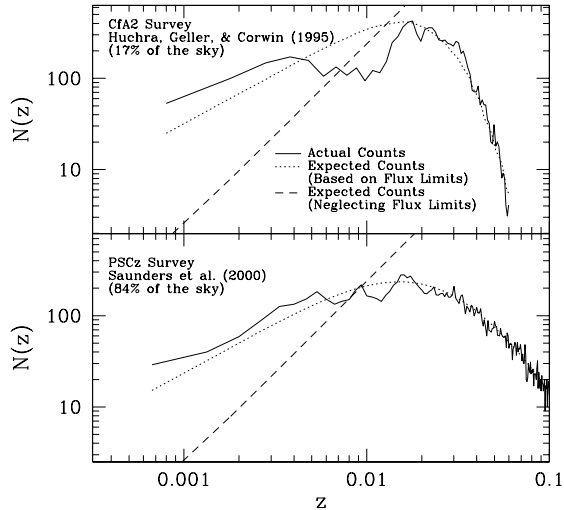,width=8cm}}
  \caption{\em {
Comparison of observed counts ({\it solid
line}) to those predicted based on the flux limits ({\it dotted line})
and those predicted neglecting the flux limits ({\it dashed
line}). The CfA2 survey is shown at top, the PSC$z$ at bottom. Both
show a local overdensity of only about a factor of two when the flux
limits are properly accounted for.
}}
 \end{center}
\end{figure}

If the flux limits are ignored, as in \cite{tanco}, the 
incorrect conclusion that we live in a large overdensity is 
easily recovered (dashed lines in Fig. 2).

As mentioned above, the CfA2 survey covers a relatively small fraction
of the sky. Thus, the PSC$z$ redshift survey provides a more useful
sample to use in the context of this paper. Using the selection
function provided by \cite{saunders00a}, we again show the expected
versus the observed counts for the PSC$z$ survey in the bottom panel
of Figure 2. This survey also shows we are living in a
slight overdensity, and furthermore reveals the general homogeneity of
the nearby universe. (The actual counts and their dependence on
redshift are slightly different than for CfA2, because the galaxies
are selected in different ways).

\section*{Calculation of the diffuse flux of UHECRs}

We calculate the diffuse flux of UHECR protons numerically 
and compare our results with the analytical calculations
carried out as in \cite{bg88}. 
Our propagation code includes pair production and photopion
production as energy losses and also adiabatic energy losses 
due to the expansion of the universe. Since the inelasticity
for pair production is very low, we consider it as a continuous
energy loss process. The magnetic field is not included.
A more detailed description of the numerical
approach and of the results is reported in \cite{bbo00}.

In order to compare the results of the simulation with the 
observed statistics of events of operating or planned detectors,
we generate the events in such a way that the total number of
events above $10^{19}$ eV equals the observed number for the specific
experiment under consideration. 

The results of the code are checked versus the analytical results for the
modification factor from single sources and from a diffuse distribution of
sources. The agreement is excellent, and the effect of the fluctuations 
at energies larger than $\sim (3-4)\times 10^{19}$ eV is evident (see below).
On average the simulated flux is slightly larger than the analytical one, as
expected on the basis of the stochasticity of the process of photopion
production (on small distances there is an appreciable chance that some
protons do not interact at all). 

We investigate the effects of the {\it real} distribution of sources
on the observed spectra of UHECRs, for different choices of the injection 
spectrum. 

The possibility that a local overdensity of sources of UHECRs may help in
solving the problem of the existence of events above the GZK cutoff goes
back to \cite{beregri79} and is summarized in \cite{bible}.
The effect can easily be understood, since
the severe photopion energy losses limit the maximum distance of UHECRs to
distances of a few Mpc, while lower energy CRs can come from much larger
distances. A local overdensity mainly affects, as a consequence, the fluxes
of UHECRs. This is illustrated in Fig. 3 for a toy model in which the
local overdensity is a top-hat function with $\Delta \rho/\rho=1,~10,~30$
(solid, dashed and dash-dotted lines respectively) in a region of 
$\sim 20$ Mpc around the Earth. The fluxes of UHECRs have been calculated 
following the analytical approach of \cite{bg88}.

\begin{figure}[thb]
 \begin{center}
  \mbox{\epsfig{file=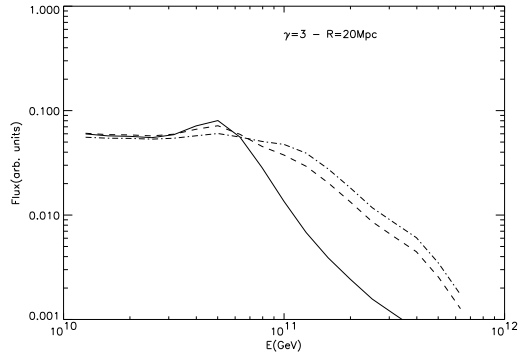,width=7cm}}
  \caption{\em {The effect of a local overdensity within $20$ Mpc on the
fluxes of UHECRs.
}}
 \end{center}
\end{figure}

In order to have a direct comparison with the results of Ref. \cite{tanco},
we first consider the case of an injection spectrum $E^{-\gamma}$ with 
$\gamma=3$ and a source distribution extended up to a maximum redshift 
$z_{max}=0.1$, that corresponds approximately to the maximum redshift
available in the PSCz catalog. 
It is worth stressing that the reason for the choice $\gamma=3$ in 
\cite{tanco} was motivated by the need to first reproduce the
results of Yoshida and Teshima \cite{yoshida}, whose curve is usually 
superimposed to the AGASA results \cite{AGASA}. However, that curve seems to
be obtained for $\gamma=2.3$ (Teshima, personal communication). 
The results of our calculations and of Ref. \cite{tanco} seem to
be incompatible with such a choice of the power index. 

We carry out the simulation for a homogeneous
distribution of the sources and then for a source distribution that 
follows the profile found in the previous section, after the correction 
for selection effects. 

\begin{figure}[thb]
\label{figure3}
 \begin{center}
  \mbox{\epsfig{file=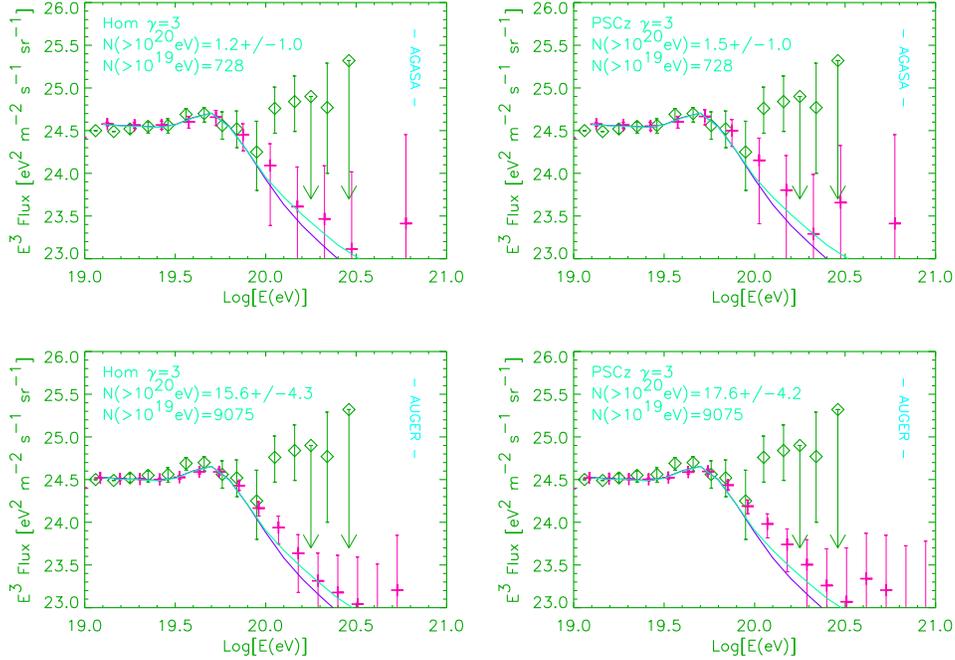,width=13cm}}
  \caption{\em {Results of the simulations for an injection spectrum 
$E^{-3}$. In the four panels the dark and light continuous lines are the
result of the analytical calculations for the homogeneous distribution of
sources and for a distribution following the PSCz catalog. The diamonds
are tha AGASA data. Upper left:
homogeneous distribution and 
$N(>10^{19}eV)$=728; upper right: PSCz catalog and $N(>10^{19}eV)$=728; 
lower left: homogeneous distribution and $N(>10^{19}eV)$=9075; 
lower right: PSCz catalog and $N(>10^{19}eV)$=9075.
}}
 \end{center}
\end{figure}
The generation of events is ended when 
the total number of events with energy above some threshold
equals the observed number. In fig. 4 we normalize the flux at
$10^{19}$ eV and we stop the generation of new events when the total number 
of events above $10^{19}$ eV becomes 728 \footnote{
This sample includes the newly released AGASA data \cite{AGASAnew}},
equal to the number of events 
detected by AGASA in that energy range. The left upper panel is for a
homogeneous distribution of the sources, while the right upper panel is
obtained by adopting the PSCz distribution of galaxies. The diamonds
are the AGASA data points and the crosses are the results of the simulation. 
The error bars in the simulation are obtained by generating 100 realizations
and calculating their mean and variance. The continuous curves represent the
result of the analytical calculation for the same value of the parameters.
The lower curve is for the homogeneous case and the upper curve for the
distribution derived from the  
PSCz catalog. One can easily see that the difference is small,
which is expected since the correction for selection effects considerably 
reduces the local overdensity in comparison to what found by 
\cite{tanco}. The total number of events with $E>10^{20}$ eV is
$1.2\pm 1.0$ for the homogeneous case and $1.5\pm 1.0$ for the PSCz catalog
galaxies. 
No one of our realizations had a number of events above $10^{20}$ eV 
comparable with observations. The main reason for that is that the 
spectrum is quite steep, in addition to the suppression due to photopion 
production.

In order to study the effect of an increased statistics of events, we 
simulated the situation for the case of Auger (9075 events predicted above
$10^{19}$ eV in the first 3 years of operation). The situation is illustrated 
in the two lower panels of fig. 4 (on the left the homogeneous
case and on the right the PSCz case). In general, the
size of the error bars decreases everywhere. The number of events at
$E>10^{20}$ eV (see figures) in the two cases is still much smaller than 
the projected one ($\sim 100$).

Since the main reason for the small number of events at high energy in the 
simulations is the steep spectra adopted above and in \cite{tanco}, 
it is natural to look for different choices. 

%In Fig. 3 we plotted our results for $\gamma=2.7$. The curves and 
%data points have the same meaning as in fig. 2. As above, the panels 
%{\it a,b)} and {\it c,d)} are respectively for the AGASA and the 
%AUGER statistics. The number of events at $E>10^{20}$ eV is still 
%smaller than the observed number in AGASA.

%\begin{figure}[thb]
% \begin{center}
%  \mbox{\epsfig{file=figure27.ps,width=15cm}}
%  \caption{\em {Same as Fig. 2 but with an injection spectrum $E^{-2.7}$.
%}}
% \end{center}
%\end{figure}

Many acceleration mechanisms produce spectra which are quite flatter 
than $E^{-3}$. We calculate the expected flux of UHECRs for an 
injection spectrum $E^{-2.1}$. In this case the normalization is chosen in
such a way that the total number of events with energy $E>4\times 10^{19}$ eV
is the same as observed by AGASA (49 events). The reason for this normalization
will be clear below. 

The results of the simulation and the corresponding analytical result are
plotted in fig. 5 for three cases: 
homogeneous distribution of the sources
with $z_{max}=0.1$ (crosses), PSCz distribution with $z_{max}=0.1$ (stars), 
homogeneous distribution with $z_{max}=1$ (squares).

\begin{figure}[thb]
\label{figure21}
 \begin{center}
  \mbox{\epsfig{file=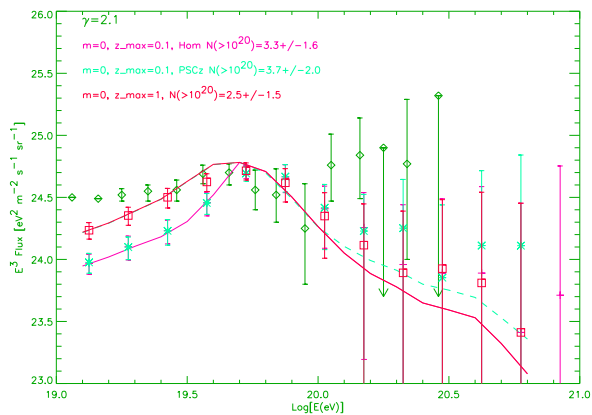,width=8cm}}
  \caption{\em {UHECRs from an injection spectrum $E^{-2.1}$. The diamonds are
the AGASA data. The results of the simulation are for an homogeneous
distribution and $z_{max}=0.1$ (crosses), for the PSCz sources and
$z_{max}=0.1$ (stars) and for an homogeneous distribution and $z_{max}=1$
(squares). The continuous lines are the analytical results for the same 
cases.
}}
 \end{center}
\end{figure}

The number of events at $E>10^{20}$ eV is mainly determined by the local 
distribution of the sources. For the adopted normalization, the homogeneous
distribution gives $3.3\pm 1.6$ events above $10^{20}$ eV (to be compared with
8) and the PSCz distribution provides $3.7\pm 2.0$ events in the same range. 
In this last case, about $5\%$ of our realizations give a number of events
above $10^{20}$ eV which is equal to or larger than the observed one.
\footnote{If the old AGASA statistics of 47 events above $4\times 10^{19}$
eV is used, $15\%$ of the realizations give a number of events above
$10^{20}$ eV equal to or larger than the 6 observed then.}
The conclusion that the number of events above $10^{20}$ eV is within 
$\sim 2$ sigmas of the observations
for an injection spectrum $\sim E^{-2}$ is consistent with
the findings in \cite{bachall}.

The deficit of events at energies lower than $\sim (3-4)\times 10^{19}$ eV 
is evident in the case $z_{max}=0.1$. The high redshift sources only 
contribute additional flux at the low energies, therefore we also considered
the case $z_{max}=1$. From fig. 5 it appears that the deficit is now evident
only at energies lower than $\sim 2\times 10^{19}$ eV, where  
additional factors might further improve the low energy agreement. 
Some possible factors are: 1) cosmological magnetic fields;
2) source evolution; 3) a separate (possibly galactic) component which 
is relevant at lower energies and has a steep spectrum. 
These possibilities are discussed at length in \cite{bbo00}.

\section*{Super Heavy Dark Matter: the direct connection}

In this section we consider another example of a possible connection between
the problem of the dark matter and the explanation of the UHECRs. 
 
Heavy particles ($m_X\sim 10^{12}-10^{14}$ GeV) can be produced in the 
Early Universe in different ways
\cite{kolb,bere,kuzmin,kt} and their lifetime can be finite though very
long compared to the present age of the universe. In these circumstances,
super-heavy particles can represent an appreciable fraction, if not all
of the cold dark matter in the universe \cite{kolb,bere,kuzmin,kt}.
The occasional decay of these
particles results in the production of UHECRs, as  
widely discussed in the literature \cite{bere,sarkar,bbv,blasi}. In particular,
if the relics cluster in 
galactic halos, as is expected, they can explain the cosmic ray 
observations above $\sim 5\times 10^{19}$ eV.  

The decay of heavy relics results usually in the production of a 
quark-antiquark pair which rapidly hadronizes, generating two jets 
with approximately $95\%$ of the energy in pions, and $\sim 5\%$ in 
baryons. The decay of the pions results in the observed high energy 
particles, mainly in the form of gamma rays, and
in the generation of ultra-high energy neutrinos. The spectrum of the gamma
photons is relatively flat ($\sim E^{-1.5}$) reflecting the behaviour of the
fragmentation function for the quarks. Therefore two main signatures of this
model are: {\it i)} a flat energy spectrum; {\it ii)} composition dominated 
by gamma rays rather than by protons. Moreover, as in all top-down models, 
heavy elements are expected to be completely absent.
Several calculations of the expected fluxes have been performed and
presented in the literature. In Fig. 6 we report the results of the
calculations of Ref. \cite{blasi}. 
\begin{figure}[thb]
\label{flussi}
 \begin{center}
  \mbox{\epsfig{file=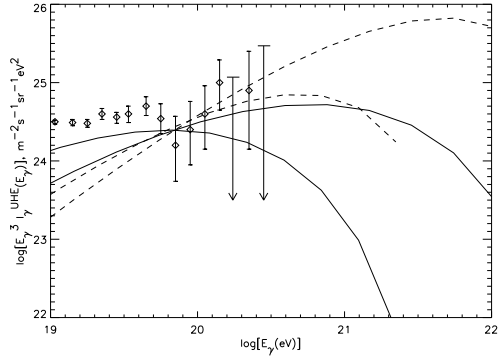,width=7cm}}
  \caption{\em {Fluxes of UHECRs from SH relics.
}}
 \end{center}
\end{figure}
The solid curves in fig. 6 are obtained by using a 
supersymmetric generalization of the quark fragmentation function, derived in 
\cite{SUSY}, while the dashed lines are obtained by using the standard
MLLA-QCD fragmentation function \cite{dok}. The thick lines are for 
$m_X=10^{14}$ GeV, while the thin lines are for $m_X=10^{13}$ GeV.

In \cite{dt,bbv} it was first recognized that, due to the asymmetric 
position of the Earth in the Galaxy, an appreciable anisotropy would
result in this model.
In \cite{bm,medina} this issue was considered
more quantitatively, taking into account the exposure of the present 
experiments.
All authors concur that the present data is consistent with 
the predictions of the relic model for practically all reasonable values  
of the model parameters.

Recently, an interesting pattern has arisen from the analysis of the 
events with  energy larger than $4\times10^{19}$ eV: in \cite{AGASA} 
the sample with this energy cut comprises 47 events \footnote{The
data in \cite{AGASAnew} were not included in this analysis.}, whose overall 
distribution in space does not show appreciable deviation from isotropy.  
However, 3 doublets and one triplet were identified within an 
angular scale of $2.5^o$, comparable with the
angular resolution of the experiment. A complete analysis, 
including the whole set of UHECR events above $4\times 10^{19}$ eV
from the existing experiments was performed in \cite{watson}. 
This extended sample comprises
92 events and shows 12 doublets and two triplets (each triplet is also 
counted as three doublets) within an angle of $3^o$.  
The chance probability of having more than this number of doublets 
was estimated to be $\sim 1.5\%$. Although it is probably too soon to 
rule out the possibility that these multiplets are just a random 
fluctuation, it is instructive to think about the possibility that 
their presence contains some physical information about the sources 
of UHECRs. Most of the top-down models for 
UHECRs (e.g. strings, necklaces, vortons, etc.) cannot naturally 
explain the multiplets. 

Here we discuss how the multiplets can be interpreted in
the context of the super-heavy dark matter (SHDM) model, following the
discussion in \cite{bs00}

We know a few things on dark matter, mainly as suggested by N-body
simulations (see for instance \cite{simul}). Dark matter seems to be clustered
in galactic halos with a distribution strongly peaked in the center.
We model this distribution as in \cite{nfw}:
\begin{equation}
n_H(r)=n^0\frac{(r/r_c)^{-1}}{\left[1+\frac{r}{r_c}\right]^2},
\label{eq:NFW}
\end{equation}
where $r_c$ is the core size and $n^0$ is a normalization parameter. 
These two parameters can be set by requiring that the halo 
contains a given total mass ($M_H$) and that the velocity dispersion 
at some distance from the center is known (in the case of the Galaxy, 
the velocity dispersion is $\sim 200$ km/s in the vicinity of our 
solar system.). 
Alternative fits to the simulated dark matter halos and a discussion of 
whether or not simulated halos appear to be consistent with observations 
are provided in \cite{simul}.

In addition to the smooth dark matter distribution, represented by
eq. (\ref{eq:NFW}), N-body simulations also show that there is a 
clumped component which contains $\sim 10-20\%$ of the total mass. 
The presence of these clumps are a natural consequence of the way in 
which gravity assembles dense virialized halos such as our galaxy 
today from the initially smooth density fluctuation field which 
was present when the cosmic microwave background (CMB) 
decoupled from the baryons. A more extended discussion on the
formation and merging of the clumps can be found in \cite{bs00} and references
therein.

We found that a good fit to 
the joint distribution in clump mass and position in the simulations 
of \cite{simul} is given by
\begin{equation}
n_{cl}(r,m) = n_{cl}^0 \left(\frac{m}{M_H}\right)^{-\alpha} 
\left[1+\left(\frac{r}{r_c^{cl}}\right)^2\right]^{-3/2},
\label{eq:clumps}
\end{equation}
where $n_{cl}^0$ is a normalization constant, $r_c^{cl}$ is the 
core of the clumps distribution, and $\alpha$ describes the relative 
numbers of massive to less massive clumps. The simulations suggest that 
$\alpha\sim 1.9$ \cite{simul}.  The constraints on the core size are 
weaker---we will study the range where $r_c^{cl}$ is between 3 and 
30 percent of $R_H$.  
In \cite{simul}, a halo with $M_H\approx 2\times 10^{12}~M_\odot$
contains about $500$ clumps with mass larger than $\sim 10^8~M_\odot$. 
This sets the normalization constant in eq. (\ref{eq:clumps}).

Clumps in the parent NFW halo are truncated at their tidal radii.  
The tidal radius of a clump depends on the clump mass, the density 
profile within the clump, and on how closely to the halo center it 
may have been.  
We assume that clumps of all mass are isothermal spheres (even though 
they are not truly isothermal, \cite{simul} suggest this is reasonably 
accurate):  $\rho_{cl}(r_{cl})\propto 1/r_{cl}^2$, 
where $r_{cl}$ is the radial coordinate measured from the center of 
the clump. 

As shown in \cite{bbv,medina}, the total (energy integrated) flux of 
UHECRs per unit solid angle from a smooth distribution of dark matter 
particles in the halo is
$\frac{d\Phi}{d\Omega} \propto \int_0^{R_{max}} dR\, n_H(r(R))$,
where $R$ is the distance from the detector, and $r$ is the distance 
from the galactic center (so $R$ and $r$ 
are related by trigonometrical relations accounting for the
off-center position of the Earth in the Galaxy). 
The upper limit, $R_{max}$, depends on the line of sight. 

It is intuitively obvious that clumped regions will give an excess 
of events from certain directions, as was first pointed out in 
\cite{bereproc}. 
Details of the numerical calculation of the fluxes and direction of arrivals 
of UHECRs can be found in \cite{bs00}.

Fig. 7 shows as an example one of the generated flux maps: 
the map represents
the ratio of the total flux including the contribution from clumps,
to the flux obtained by using a smooth NFW profile.  The various free 
parameters were $r_c=8$ kpc, $r_c^{cl}=10$ kpc, and the mass distribution 
was truncated at a clump mass of $1\%$ of the mass of the NFW halo. 
This sort of plot emphasizes the clump contribution. 

\begin{figure}[thb]
 \begin{center}
  \mbox{\epsfig{file=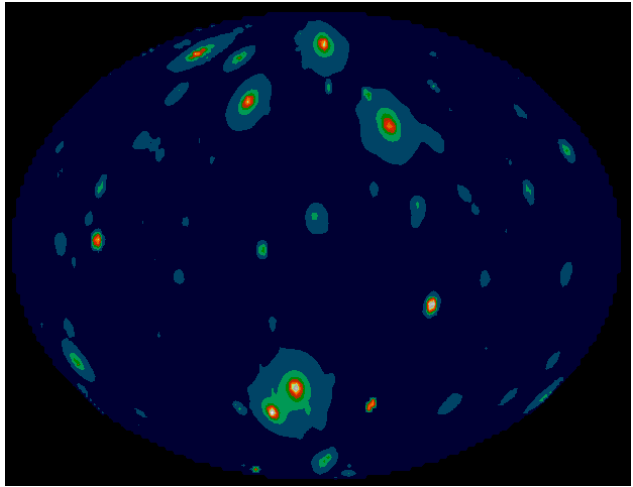,width=9cm}}
  \caption{\em {Small scale anisotropies in the model of super-heavy 
relics in the halo.
}}
 \end{center}
\end{figure}

To calculate the small scale anisotropies, we generated $10^4$ mock 
samples, each of 92 observed events, and counted the number of 
doublets and triplets for angular scales of 3, 4, and 5 degrees. 
Our codes can also be used to check the corresponding numbers for 
the case of isotropic arrival directions (as in \cite{watson}). Two
sets of values of the cores for the NFW and the clumped component
were adopted, one
in which $r_c=8$ kpc and $r_c^{cl}=10$ kpc (case 1) and the other with 
$r_c=r_c^{cl}=20$ kpc (case 2). The observed numbers of doublets within
3, 4, and 5 degrees for an isotropic distribution of arrival directions
are given in \cite{watson} and are 12, 14 and 20 respectively. 
The number of doublets that we obtain in case 1 are 8, 14, and 21
within 3, 4, and 5 degrees respectively. The probability that 
the number of doublets equals or exceeds that observed is
$12\%$, $47\%$ and $57\%$ respectively.  This should be compared with 
the $1.5\%$, $13.4\%$ and $15.9\%$ quoted in \cite{watson} for an
isotropic distribution of arrival directions. 

We repeated the same calculation for the case 2. The corresponding 
averages and probabilities of exceeding the observed number of 
doublets within 3, 4 and 5 degree scales are 6.6, 12, and 18, 
and $4.5\%$, $29\%$ and $36\%$ respectively. 

In both cases 1 and 2, the number of doublets on angular scales of 
4 and 5 degrees is consistent with the observed values; presumably 
the discrepancy at 3 degrees is random chance. 

We have also studied the occurrence of triplets.  
There is some ambiguity as to how a triplet is best defined; we have 
chosen to define triplets as configurations in which all three pairs 
would have been classified as doublets.  (This means, for example, that 
a co-linear configuration of two doublets is not necessarily a triplet.)   
With this definition, the average number of triplets in case one is 
0.5, 1.5 and 3, with the probability of having more than the 
observed triplets (2, 2, 3 respectively) equal to 4\%, 16\% and 35\%.  
For case 2, the correspondent numbers are 0.4, 1, and 2.5 triplets 
and $2\%$, $8\%$ and $20\%$ for the probabilities to have more triplets
than observed.

It is instructive to explore the reasons for the multiple events 
in the SHDM model.
If we study the case in which all the halo mass is in the smooth 
NFW component, then the number of doublets typically drops by 
one or two.  This suggests that the anisotropy due to our position 
in an NFW halo can result in a number of multiplets of events which 
is considerably larger than if the arrivals were from an isotropic 
background.  The number of multiplets from the clumped component is 
mainly affected by the presence of large nearby clumps, whose number 
depends on the high mass cutoff imposed in the mass function of clumps. 
A maximum mass of $1\%$ of the halo mass implies a total mass in the 
clumps of $\sim 10-15\%$ of $M_H$, 
consistent with the results of the simulations \cite{simul}. 
Larger cutoffs imply larger mass fractions, 
which are harder to reconcile with the N-body simulations.

\section*{Conclusions}

We discussed here two possible connections between the problem of origin and
clustering properties of dark matter and the nature of the particles with 
energy in excess of $10^{20}$ eV. 

In the first part we investigated an indirect connection, consistent
in the effect on the fluxes of UHECRs due to local clustering of the
sources, driven by the corresponding clustering of dark matter.
It has been known for a long time \cite{beregri79,bible} that a local
overdensity in the sources of UHECRs may shift the energy of the GZK 
cutoff towards larger energies. The overdensities 
needed to make this effect relevant as a possible solution of the
UHECR puzzle are of the order of $\sim 10$ or more. In a recent paper,
Medina-Tanco \cite{tanco} claimed that this is exactly the value extracted 
from large scale structure surveys like the CfA2 catalog. We reconsidered this
problem and showed that, by correctly extracting the density field from the
CfA2 and the PSCz catalogs, the local overdensity is not larger than $\sim 2$,
so that the calculations of the propagation of UHECRs are not
particularly affected by the local source distribution. In other words the
increase in the number of events above $10^{20}$ eV, in comparison with the
case of homogeneous distribution is not important \cite{bbo00}.

On the other hand we noticed an interesting point: using an injection
spectrum $E^{-2.1}$ typical of astrophysical sources, we obtain a number 
of events with energy above $10^{20}$ eV, which is basically compatible with
the observed number, once the statistical fluctuations and the Poisson 
noise in the photopion production have been taken into account. The agreement
of the prediction at lower energies, depend on additional physics 
(evolution of the sources, cosmological magnetic fields or an additional 
(possibly) galactic component at energies around $(1-2)\times 10^{19}$ eV)
that was not included in our calculations.

In the second part of the paper, we investigated a direct connection between 
dark matter and UHECRs. It has been proposed that most of the dark matter 
might be made of super heavy quasi-stable particles, created in the early 
universe. These particles would cluster in large scale structures, and in
particular in the galactic halo. The rare decays of the super heavy dark
matter particles in the halo can easily explain the fluxes of UHECRs that 
are observed. We reviewed the testable predictions of this model and then 
concentrated on a particular aspect, the anisotropy of arrival directions.
A large scale (dipole) anisotropy in this model is easily foreseeable, since
the earth is off center in the Galaxy. The predicted large scale anisotropy is 
compatible with the observed one \cite{bm,medina}, mainly due to a lack 
of exposure in the
direction of the galactic center. Based on the clustering properties of dark 
matter on smaller scales, we also discussed
the recent results of \cite{bs00} on the 
the small scale anisotropies. Based on the informations provided by N-body
simulations of structure formation, the combination of the smooth peculiar
density profile in the Galaxy and the clumped component survived in 
the halo can explain the observed small scale anisotropies, in terms of
doublets and triplets of events.
Future full sky experiments as Auger \cite{auger} will be extremely 
important to
test this model, mainly in two ways: 1) an increased statistics of
events will allow a better determination of the small angle clustering,
and 2) a better composition determination will definitely allow to 
understand if gamma rays are an important component of the UHECRs.

{\bf Aknowledgments} I am grateful to my collaborators, 
M. Blanton, A.V. Olinto and R.K. Sheth, for continuos interactions. 
I am also grateful to the organizers of the 
``International Workshop on Observing UHECRs from Space and Earth''
for the nice environment that allowed instructive discussions
with P. Biermann, J. Ellis, T. Gaisser, D. Harari, A. Letessier-Selvon, 
J. Linsley, L. Scarsi, G. Sigl, M. Teshima and T. Weyler among 
others. This work was
supported by the DOE and the NASA grant NAG 5-7092 at Fermilab.


\begin{references}

\bibitem{greis}
Greisen, K., {\it Phys. Rev. Lett.} {\bf 16}, 748 (1966).

\bibitem{kz}
Zatsepin, G. T., and Kuzmin, V. A., {\it Pis'ma Zh. Ekps. Teor. Fiz.}
{\bf 4}, 114  (1966) [{\it JETP Lett.} {\bf 4}, 78 (1966)].

\bibitem{beregri79}
Berezinsky, V. S., and Grigorieva, S. I., {\it Proc. 16th. Int. Cosmic
Ray Conf., Kyoto} {\bf 2}, 81 (1979).

\bibitem{kolb}
Chung, D. J. H., Kolb, E. W., and Riotto, A., {\it Phys. Rev. Lett.} 
{\bf 81}, 4048 (1998).

\bibitem{bere}
Berezinsky, V. S., Kachelriess, M., and Vilenkin, A., {\it Phys. Rev. 
Lett.} {\bf 79} 4302 (1997).

\bibitem{kuzmin}
Kuzmin, V. A., presented at the Workshop ``Beyond the Desert'', Castle 
Ringberg, 1997, astro-ph/9709187; International Workshop on 
Non-Accelerator New Physics, Dubna, 1997.

\bibitem{kt}
Kuzmin, V. A. and Tkachev, I. I., preprint astro-ph/9903542.

\bibitem{bbo00}
Blanton, M., Blasi, P., and Olinto, A. V., in preparation.

\bibitem{davis82a}
Davis, M., \& Huchra, J., {\it Astrophys. J.} 254, 437 (1982).

\bibitem{strauss95a} 
Strauss, M.~A., \& Willick, J.A., {\it Phys. Rep.} {261}, 271 (1995).

\bibitem{tanco}
Medina Tanco, G. A., {\it Astrophys. J.} {\bf 510}, 91 (1999).

\bibitem{huchra95a}
Huchra, J.~P., Geller, M.~J., \& Corwin, Jr., H.~G., {\it Astrophys. J. 
Suppl.} 70, 687 (1995).

\bibitem{saunders00a}
% PSCz paper
Saunders, W., Sutherland, W.~J., Maddox, S.~J., Keeble, O., Oliver,
S.~J., Rowan-Robinson, M., McMahon, R.~G., Efstathiou, G., Tadros, H.,
White, S.~D.~M., Frenk, C.~S., Carraminana, A., Hawkins,
M.~R.~S., submitted to {\it MNRAS}, preprint (astro-ph/0001117).

\bibitem{peebles80a}
Peebles, P.~J.~E., The Large-Scale Structure of the
Universe (Princeton, NJ: Princeton University Press) 1990.

\bibitem{efstathiou88a}
Efstathiou, G., Ellis, R.~S., \& Peterson, B.~S., {\it MNRAS} 232, 
431 (1988).

\bibitem{sandage79a}
% STY luminosity function fitting method
Sandage, A., Tammann, G.~A., \& Yahil, A., {\it Astrophys. J.} 232, 352 
(1979).

\bibitem{bg88}
Berezinsky, V. S., and Grigorieva, S. I., {\it Astron. Atroph.} {\bf 199},
1 (1988).

\bibitem{bible}
Berezinsky, V. S., et al. {\it Astrophysics of cosmic rays}, 
Amsterdam: North-Holland, 1990, edited by Ginzburg, V.L.

\bibitem{yoshida}
Yoshida, S., and Teshima, M., {\it Prog. Theor. Phys.} {\bf 89}, 833 (1993).

\bibitem{AGASA}
Takeda, M., et al., {\it Phys. Rev. Lett.} {\bf 81}, 1163 (1998).

\bibitem{AGASAnew}
Hayashida, N., et al., Appendix to {\it Astrophys. J.} {\bf 522}, 225 (1999)
(preprint astro-ph/0008102).

\bibitem{bachall}
Bahcall, J. N., and Waxman, E., to appear in {\it Astrophys. J.} 
(preprint hep-ph/9912326).

\bibitem{sarkar}
Birkel, M., and Sarkar, S., {\it Astropart. Phys.} {\bf 9}, 297 (1998).

\bibitem{bbv}
Berezinsky, V. S., Blasi, P., and Vilenkin, A., {\it Phys. Rev.} 
D{\bf 58}, 103515 (1998).

\bibitem{blasi}
Blasi, P., {\it Phys. Rev.} D{\bf 60}, 023514 (1999).

\bibitem{SUSY}
Berezinsky, V. S., and Kachelriess, M., {\it Phys. Lett.} B{\bf 434}, 61
(1998).

\bibitem{dok}
Dokshitzer, Yu. L., et al., {\it Basics of Perturbative QCD}, Editiones
Fronti\`eres, Gif-sur-Yvette, France, 1991.

\bibitem{dt} 
Dubovsky, S. L., and Tynyakov, P. G., {\it Pis'ma Zh. Eksp. Teor. Fiz.} 
{\bf 68}, 99 (1998) [{\it JETP Lett.} {\bf 68}, 107 (1998)].

\bibitem{bm}
Berezinsky V. S., and Mikhailov, A., {\it Phys. Lett.} B{\bf 449}, 237 
(1999).

\bibitem{medina}
Medina Tanco G. A., and Watson, A. A., {\it Astropart. Phys.} {\bf 12}, 
25 (1999).

\bibitem{watson}
Uchihori, Y., et al., {\it Astropart. Phys.} {\bf 13}, 151 (2000).

\bibitem{bs00}
Blasi, P., and Sheth, R. K., {\it Phys. Lett.} B{\bf 486}, 233 (2000).

\bibitem{simul}
Ghigna, S., Moore, B., Governato, F., Lake, G., Quinn, T., Stadel, J.,
preprint astro-ph/9910166; Moore, B., Ghigna, S., Governato, F.,  
Lake, G., Quinn, T., Stadel, J., Tozzi, P., {\it Astrophys. J.} {\bf 524}, 
19 (1999); Tormen, G., Diaferio, A., Syer, D., {\it MNRAS} {\bf 299}, 
728 (1998).

\bibitem{nfw}
Navarro, J. F., Frenk, C. S., and White, S. D. M., {\it Astrophys. J.} 
{\bf 462}, 563 (1996).

\bibitem{bereproc}
Berezinsky, V., Invited talk at TAUP-99, Paris, September 6 - 10, 1999, 
preprint hep-ph/0001163.

\bibitem{auger}
Cronin, J. W., {\it Rev. Mod. Physics} {\bf 71}, 175 (1999).

\end{references}
\end{document}